\documentclass[showpacs,preprintnumbers,amsmath,amssymb,superscriptaddress,nofootinbib]{revtex4}

\usepackage{graphicx}% Include figure files
\usepackage{bm}% bold math
\usepackage{epsfig}

\usepackage{color}

\newcommand{\matr}{\mathrm}

\allowdisplaybreaks

\begin{document}

%- {{{ Header:

\preprint{TTP16-033}
\title{Analytic three-loop static potential}

\author{Roman N. Lee}
\affiliation{Budker Institute of Nuclear Physics, 630090 Novosibirsk, Russia}

\author{Alexander V. Smirnov}
\affiliation{Research Computing Center, Moscow State University, 119992 Moscow, Russia}

\author{Vladimir A. Smirnov}
\affiliation{Skobeltsyn Institute of Nuclear Physics of Moscow State University, 119992 Moscow, Russia}

\author{Matthias Steinhauser}
\affiliation{Institut f\"ur Theoretische Teilchenphysik, Karlsruhe
  Institute of Technology (KIT), D-76128 Karlsruhe, Germany}

%\date{}

%
\begin{abstract}
We present analytic results for 
the three-loop static potential of two heavy quarks.
The analytic calculation of the missing ingredients is outlined and results
for the singlet and octet potential are provided.
\end{abstract}

\pacs{12.38.Bx, 14.65.Dw, 14.65.Fy, 14.65.Ha}

\maketitle

%- }}}

%- {{{ Intro:

\section{Introduction}

The static potential between two heavy quarks belongs to the fundamental
quantities of QCD.  In lowest order it is described by the Coulomb potential
adapted to QCD.  Such an approach has already been used more than 40 years
ago~\cite{Appelquist:1974zd} to describe the bound state of heavy quarks.
Shortly afterwards the one-loop corrections were
computed~\cite{Fischler:1977yf,Billoire:1979ih} and the two-loop terms were
added towards the end of the
nineties~\cite{Peter:1996ig,Peter:1997me,Schroder:1998vy}.  Light quark mass
effects at two loops can be found in Ref.~\cite{Melles:2000dq}. About eight
years ago the three-loop corrections have been computed by two groups in
Refs.~\cite{Smirnov:2008pn,Smirnov:2009fh,Anzai:2009tm}. However, in contrast
to the lower-order expressions, the three-loop results could only be presented
in numerical form. In fact, in Refs.~\cite{Smirnov:2008pn,Smirnov:2009fh}
three coefficients in the expansion of the master integrals around $d=4$,
where $d$ is the space-time dimension, could only be evaluated numerically
(see also below). The evaluation of one of them is described in detail in
Ref.~\cite{LSS16} (in a broader context) and the remaining two coefficients
are considered in Section~\ref{sec::P10Q10} of this paper.  We are thus in the
position to present analytic results at three loops. The corresponding
expressions can be found
in Section~\ref{sec::sing}.

A generalization of the three-loop singlet potential has been considered in
Ref.~\cite{Anzai:2010td}.  It is still assumed that the heavy colour sources
form a singlet state, however, the colour representation is kept general.

The new results can also be used to present analytic expressions for the
so-called octet potential which describes the situation where the quark and
anti-quark do not form a colour-singlet but a colour-octet state.  Two-
and (numerical) three-loop results have been obtained in
Refs.~\cite{Kniehl:2004rk,Collet:2011kq} and~\cite{Anzai:2013tja},
respectively.  Analytic results for the octet potential are
presented in Section~\ref{sec::oct}.

In order to fix the notation we write the momentum space potential in
the form
\begin{eqnarray}
  V^{[c]}(|{\vec q}\,|)&=&
  -{4\pi C^{[c]} \frac{\alpha_s(|{\vec q}\,|)}{{\vec q}\,^2}}
  \Bigg[1+\frac{ \alpha_s(|{\vec q}\,|) }{4\pi} a_1^{[c]}
    +\left( \frac{\alpha_s(|{\vec q}\,|)}{4\pi}\right)^2a_2^{[c]}
%    \nonumber\\&&\mbox{}
    +\left(\frac{\alpha_s(|{\vec q}\,|)}{4\pi}\right)^3
    \left(a_3^{[c]}+ 8\pi^2 C_A^3\ln\frac{\mu^2}{{\vec q}\,^2}\right)
    +\cdots\Bigg]\,,
  \label{eq::V}
\end{eqnarray}
with $C^{[1]}=C_F$ for the colour-singlet and $C^{[8]}=C_F-C_A/2$ for the
colour-octet case. Here, $C_A=N_c$ and $C_F=(N_c^2-1)/(2N_c)$ are the
eigenvalues of the quadratic Casimir operators of the adjoint and fundamental
representations of the SU$(N_c)$ colour gauge group, respectively.  The strong
coupling $\alpha_s$ is defined in the $\overline{\rm MS}$ scheme and for the
renormalization scale we choose $\mu=|{\vec q}\,|$ in order to suppress the
corresponding logarithms. The general results, both in momentum and coordinate
space, can, e.g., be found in Appendix~A of Ref.~\cite{Anzai:2013tja}.

The logarithmic term in Eq.~(\ref{eq::V}) has its origin in an infra-red
divergence which is present for the first time at three loops as has been
pointed out in Ref.~\cite{Appelquist:1977es}.  The corresponding pole has been
subtracted minimally. Its presence can be understood in the context of methods
of regions and potential non-relativistic
QCD~\cite{Pineda:1997bj,Beneke:1997zp,Beneke:1998jj,Kniehl:1999ud,Brambilla:1999xf}
where $V^{[c]}$ appears as a matching coefficient. Thus, the infrared
divergence cancels against ultraviolet divergences of the ultrasoft
contributions. The latter have been studied in
Refs.~\cite{Brambilla:1999qa,Kniehl:1999ud,Kniehl:2002br}. For the resummation
of leading and next-to-leading ultrasoft logarithms we refer
to~\cite{Pineda:2000gza,Brambilla:2009bi,Pineda:2011db}.

The three-loop coefficient $a_3$ only has a moderate numerical value (see,
e.g., discussion in Ref.~\cite{Smirnov:2009fh}) and has thus only a relative
small influence on phenomenological quantities. This is in contrast to the
two-loop coefficient which is of the same order of magnitude as $a_1$.
However, since the static potential is a matching coefficient, it is hence not
a physical quantity. In fact, $a_3$ is scheme dependent and only the
combination with all other building blocks leads to meaningful quantities.

For later convenience we decompose the three-loop corrections according 
to the number of closed fermion loops
\begin{eqnarray}
  a_3^{[c]} &=& a_3^{[c],(3)} n_l^3 + a_3^{[c],(2)} n_l^2 +  
  a_3^{[c],(1)} n_l +  a_3^{[c],(0)}
  \,,
\end{eqnarray}
where $n_l$ is the number of light (massless) quarks.
We furthermore consider the difference between the singlet and octet
contributions and write ($i=0,1,2,3$)
\begin{eqnarray}
  a_3^{[8],(i)} &=& a_3^{[1],(i)} + \delta a_3^{[8],(i)}
  \,.
  \label{eq::delta_a3}
\end{eqnarray}
In Section~\ref{sec::oct} we provide analytical results for $\delta
a_3^{[8],(i)}$.

The three-loop coefficient of the colour singlet potential, $a_3^{[1]}$,
has entered a number of physical applications 
as building block (see also Ref.~\cite{Beneke:2015zqa} for a recent review on
applications of non-relativistic QCD to high-energy processes).  To name a few
of them we want to mention the next-to-next-to-next-to-leading order
corrections to the leptonic decay width of the $\Upsilon(1S)$
meson~\cite{Beneke:2014qea} and the top quark threshold production in electron
positron colliders~\cite{Beneke:2015kwa}.  Furthermore, $a_3$ has entered
analyses to determine precise values for the charm and bottom quark
masses~\cite{Penin:2014zaa,Ayala:2014yxa,Beneke:2014pta,Kiyo:2015ufa} and the
strong coupling constant~\cite{Bazavov:2014soa}.

%- }}}
%- {{{ $I_{11}$ and  $I_{16}$:

\section{Calculation of $I_{11}$ and  $I_{16}$\label{sec::P10Q10}}

\begin{figure}[t]
  \centering
  \begin{tabular}{cccc}
    \mbox{}\quad \includegraphics[width=.2\textwidth]{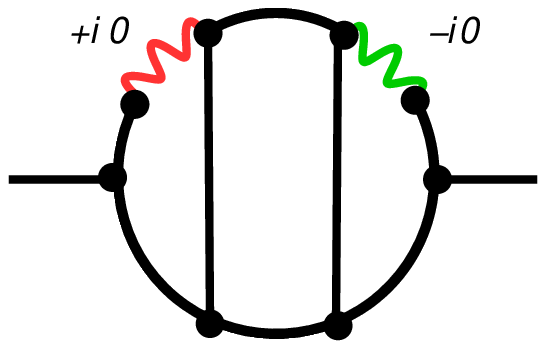} \quad \mbox{}&
    \mbox{}\quad \includegraphics[width=.2\textwidth]{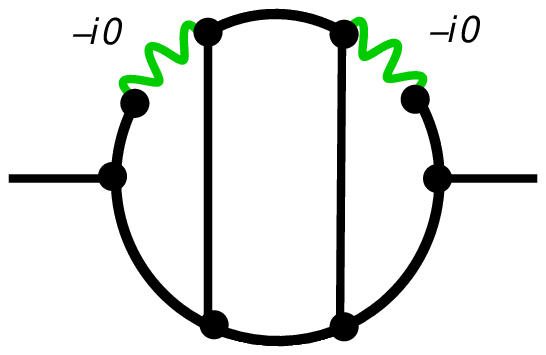} \quad \mbox{}&
    \mbox{}\quad \includegraphics[width=.2\textwidth]{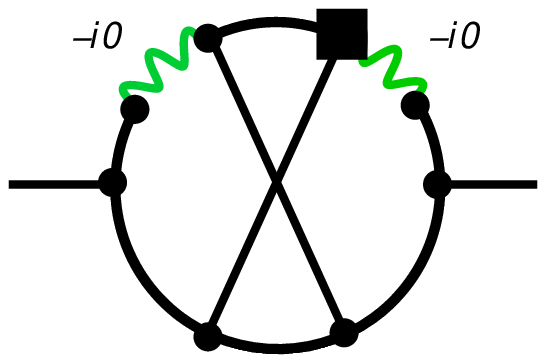} \quad \mbox{}&
    \mbox{}\quad \includegraphics[width=.2\textwidth]{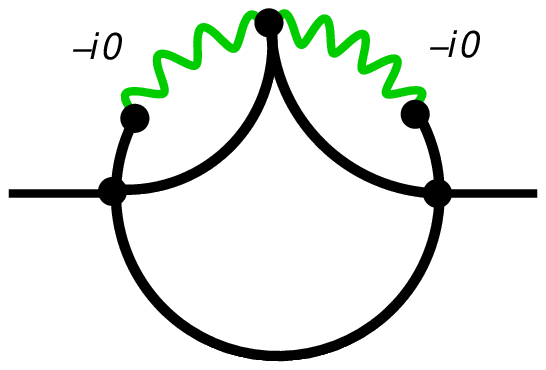} \quad \mbox{}
    \\[2em]
    (a) $I_{11}$ & (b) $I_{16}$ & (c) $I_{18}$ & (d) $I_{14}$
  \end{tabular}
  \caption{\label{fig::3MIs}(a)-(c): Master integrals entering $a_3^{[c]}$ which were
    only known numerically. Solid lines denote relativistic scalar propagators
    and wavy lines refer to static propagators. For the latter the causality
    prescription is given explicitly where $\pm i0$ indicates a propagator of
    the form $1/(-k_0 \pm i0)$ with $k_0$ being the zeroth component of the
    momentum flowing through the corresponding line. The square in $I_{18}$
    indicates a convenient choice for the numerator which is specified in
    Ref.~\cite{LSS16}. $I_{18}$ is finite and only the ${\cal O}(\epsilon^0)$
    term is needed. For $I_{11}$ and $I_{16}$ also the ${\cal O}(\epsilon^1)$
    terms enter $a_3^{[c]}$. (d): Master integral which is needed for the
    computation of the integrals in (b) and (c). The integral $I_{15}$
    belongs to the same integral family as  $I_{14}$, however, has an
    additional dot on the lower line.}
\end{figure}

The calculation of $a_3^{[1]}$ as performed in~\cite{Smirnov:2009fh} requires
the evaluation of 41 master integrals which can be sub-divided into three
different classes: There are ten integrals which do not have any static line
(i.e. a propagator of the form $1/(-k_0 \pm i0)$, see also
Fig.~\ref{fig::3MIs}), and are thus known since long. Furthermore, we have
14 integrals with a massless one-loop insertion. They can easily be integrated
in terms of $\Gamma$ functions using standard techniques. The corresponding
results have been presented in Ref.~\cite{Smirnov:2010gi}.  Results for 16
more complicated integrals can be found in Ref.~\cite{Smirnov:2010zc} as
expansions in $\epsilon=(4-d)/2$ to the necessary order except for two
integrals ($I_{11}$ and $I_{16}$ of Ref.~\cite{Smirnov:2010zc}, see also
Fig.~\ref{fig::3MIs}(a) and (b)). Their ${\cal O}(\epsilon)$ terms enter
$a_3^{[1]}$, however, they were only known numerically.  The evaluation of
these coefficients will be described in the remainder of this section. For
completeness we want to mention that the third numerical ingredient required
in~\cite{Smirnov:2009fh} comes from the finite diagram in
Fig.~\ref{fig::3MIs}(c) (the 41$^{\rm th}$ master integral) which has been
computed in a parallel article~\cite{LSS16}.

Let us also mention that techniques which have been used to compute master
integrals in~\cite{Anzai:2009tm} can be found in Ref.~\cite{Anzai:2012xw}, see
also~\cite{Sumino:2016sxe} for a status report of the
approach used in Ref.~\cite{Anzai:2009tm}.

The method which is used to compute $I_{11}$ and $I_{16}$ is based on the
dimensional recurrence relation and analyticity with respect to space-time
dimensionality $d$ (the so-called ``DRA method'') and has been developed in
Ref.~\cite{Lee:2009dh}.  In Ref.~\cite{Lee:2012te} this method has been
applied for the first time to the case with more than one master
integral in a sector.  Some integrals taken from families of integrals for the
three-loop static quark potential and denoted in~\cite{Lee:2012te}
by $I_{14}$ and $I_{15}$ (see Fig.~\ref{fig::3MIs}(d)) have been calculated.
Note that $I_{14}$ and $I_{15}$ are the only nontrivial integrals entering
the right-hand side of the dimensional recurrence relation for
$I_{16}$. Therefore, in principle, the results of Ref.~\cite{Lee:2012te} make the
calculation of $I_{16}$ straightforward.
 
However, the numerical issues related to the calculation of contributions
to the inhomogeneous terms proportional to $I_{14}$ and $I_{15}$ in the
right-hand side of dimensional recurrence relations for $I_{16}$ are
quite involved. The most complicated part of this contribution has the form
\begin{equation}
  T(\nu)=\sum_{k=0}^\infty v^T(\nu+k)\sum_{n=k}^\infty
  \left(\prod_{l=k}^{n}\matr{M}(\nu+l)\right)u(\nu+n)\,, 
  \label{eq::T}
\end{equation}
where $\nu=d/2$, $v^T(x)$, $\matr{M}(x)$, and $u(x)$ are a row-vector,
a $2\times2$ matrix and a column-vector, respectively. Their components
are rational functions of the variable $x$. In order to calculate the
sums in Eq.~(\ref{eq::T}) without nested loops, we apply the standard trick
of the DRA method, see Ref. \cite{Lee:2012hp}. Namely, let us denote
\begin{equation}
  F(k)=\sum_{n=k}^\infty \matr{P}(k,n)u(\nu+n)\,, 
  \label{eq:FasSum}
\end{equation}
where $\matr{P}(k,n)=\prod_{l=k}^{n}\matr{M}(\nu+l)$. Then 
\begin{equation}
  T(\nu)=\sum_{k=0}^\infty v^T(\nu+k)F(k)\,.\label{eq:TviaF}
\end{equation}
Using Eq.~\eqref{eq:FasSum}, the function $F(k)$ can be calculated for given
$k$ in one loop if one takes into account the recurrence relation
$\matr{P}(k,n+1)=\matr{P}(k,n)\matr{M}(\nu+n+1)$.  Now we note that $F(k)$
satisfies the recurrence relation
\begin{equation}\label{eq:Frr}
  F(k+1)=\matr{M}^{-1}(\nu+k)F(k)-u(\nu+k)\,.
\end{equation}
Therefore, in order to calculate consecutive terms of the sum in
Eq.~\eqref{eq:TviaF} we need to use Eq.~\eqref{eq:FasSum} only once, and then
use the recurrence relation \eqref{eq:Frr}. However, the price we have to pay
is much higher than for scalar sums. This is connected with the multiplication
by the inverse matrix $\matr{M}^{-1}(\nu+k)$. For $x\to \infty$ the elements of
$\matr{M}(x)$ are of order unity, while its determinant tends to
$1/1024$. Due to this fact, the multiplication by $\matr{M}^{-1}$ involves
large cancellations which result in rapid precision loss.
For example, using a precision of 7000 digits in the initial expression
we obtain only about 370 digits in the final result.

Besides, it appears that the sum over $n$ in the definition of $F(k)$
converges very slowly, with the summand behaving as $n^{-\alpha}$
($\alpha>1$) at large $n$. So, in order to obtain the high-precision
numerical result suitable for using PSLQ~\cite{PSLQ}, one has to apply the
matrix analog of the convergence acceleration algorithm described in
Ref.~\cite{LeeMingulov2016}.  In particular, one needs to know the exponent
$\alpha$ of the power-like decay. This appears to be possible thanks to
Ref.~\cite{Tulyakov2011}, where a method for finding the asymptotic behaviour
of the solutions of recurrence relations was developed. Once we dealt with
these numerical issues, we have obtained the result\footnote{See
  Fig.~\ref{fig::3MIs}(b) for a graphical definition and Eq.~(4.1)
  of Ref.~\cite{Lee:2012te} normalization factors.}
\begin{eqnarray}
I_{16}&=&
-\frac{56 \pi ^4}{135 \epsilon }
-\left(\frac{112 \pi ^4}{135}
+\frac{16 \pi ^2 \zeta (3)}{9}
+\frac{8 \zeta (5)}{3}\right)
+\left(\frac{968 \zeta (5)}{3}
-16 \pi ^4 l_2
+\frac{136 \zeta(3)^2}{3}
+\frac{400 \pi ^2 \zeta (3)}{9}
-\frac{838 \pi ^6}{2835}
  \nonumber\right.\\&&\left.\mbox{}
+\frac{1792 \pi ^4}{135}\right) \epsilon
+\bigg(\frac{6144 s_6 l_2}{7}
-\frac{6144 s_{\text{7a}}}{7}
+\frac{15360 s_{\text{7b}}}{7}
+1536 \alpha _4 \zeta(3)
+1024 \pi ^2 \alpha _5
-256 \pi ^2 \alpha _4
-\frac{64}{9} \pi ^4 l_2^3
\nonumber\\&&\mbox{}
-2976 \zeta (5)   l_2^2
-64 \pi ^2 \zeta (3) l_2^2
-\frac{112}{3} \pi ^4 l_2^2
-\frac{7680 \zeta (3)^2 l_2}{7}
-\frac{544 \pi ^6 l_2}{315}
+128 \pi ^4 l_2
+\frac{306202 \zeta (7)}{21}
-\frac{12182 \pi ^2 \zeta (5)}{7}
\nonumber\\&&\mbox{}
+\frac{64 \zeta (5)}{3}
-\frac{1168 \zeta (3)^2}{3}
-\frac{11828 \pi ^4 \zeta (3)}{945}
+\frac{1664 \pi ^2 \zeta (3)}{9}
+\frac{1376 \pi ^6}{135}
-\frac{12544 \pi ^4}{135}
+768s_6\bigg) \epsilon ^2+O\left(\epsilon ^3\right)
\,,
\label{eq::I16}
\end{eqnarray}
where $\zeta(n)$ is Riemann's zeta function evaluated at $n$ and
\begin{eqnarray}
  l_2 &=& \log(2)\,, \nonumber\\
  \alpha_n&=&\mathrm{Li}_n(1/2)+\frac{(-\log 2)^n}{n!}\,, \nonumber\\
  s_6 &=& \zeta(-5,-1)+\zeta(6)\,, \nonumber\\
  s_{7a} &=& \zeta(-5,1,1)+\zeta(-6,1)+\zeta(-5,2)+\zeta(-7) \,, \nonumber\\
  s_{7b} &=& \zeta(7)+\zeta(5,2)+\zeta(-6,-1)+\zeta(5,-1,-1) \,.
\end{eqnarray}
$\zeta(m_{1},\dots,m_{k})$ are multiple zeta values given by
\begin{eqnarray}
  \zeta(m_{1},\dots,m_{k}) &=&
  \sum\limits _{i_{1}=1}^{\infty}\sum\limits
  _{i_{2}=1}^{i_{1}-1}\dots\sum\limits _{i_{k}=1}^{i_{k-1}-1}\prod\limits
  _{j=1}^{k}\frac{\mbox{sgn}(m_{j})^{i_{j}}}{i_{j}^{|m_{j}|}}
  \,. 
\end{eqnarray}

In order to apply the DRA method to $I_{11}$, one has to take into account
that the dimensional recurrence relation for $I_{11}$ contains now two
non-trivial integrals denoted in~\cite{Lee:2012te} by $I_{9}$ and
$I_{10}$. So, in a first step one has to apply the DRA method to these two
integrals. Fortunately, they can be calculated along the same lines as
$I_{14}$ and $I_{15}$ from which they differ only by the $\pm 
i0$ prescription in one of the linear denominators.  In particular, the
summing factor has the same form as in Ref.~\cite{Lee:2012te} (see Eq. (4.14)
of that paper). Plugging the results for $I_{9}$ and $I_{10}$ in
the dimensional recurrence relation for $I_{11}$ and applying the DRA method,
we obtain\footnote{See
  Fig.~\ref{fig::3MIs}(a) for a graphical definition and Eq.~(4.1)
  of Ref.~\cite{Lee:2012te} normalization factors.}
\begin{eqnarray}
I_{11}&=&
\frac{64 \pi ^4}{135 \epsilon }
+\left(\frac{128 \pi ^4}{135}
+\frac{32 \pi ^2 \zeta (3)}{9}
-\frac{8 \zeta (5)}{3}\right)
+\left(16 \pi ^4 l_2
+\frac{968 \zeta (5)}{3}
+\frac{136 \zeta (3)^2}{3}
-\frac{800 \pi ^2 \zeta (3)}{9}
+\frac{548 \pi ^6}{2835}
-\frac{2048 \pi ^4}{135}\right) \epsilon
  \nonumber\\&&\mbox{}
+\bigg(\frac{6144 s_6 l_2}{7}
-\frac{6144 s_{\text{7a}}}{7}
+\frac{15360 s_{\text{7b}}}{7}
+1536 \alpha _4 \zeta (3)
-2048 \pi ^2 \alpha _5
+512 \pi ^2 \alpha _4
-\frac{64}{9} \pi ^4 l_2^3
-2976 \zeta (5) l_2^2
  \nonumber\\&&\mbox{}
-64 \pi ^2 \zeta (3) l_2^2
+\frac{80}{3} \pi ^4 l_2^2
-\frac{7680 \zeta (3)^2 l_2}{7}
-\frac{208 \pi ^6 l_2}{315}
-128 \pi ^4 l_2
+\frac{306202 \zeta (7)}{21}
+\frac{1482 \pi ^2 \zeta (5)}{7}
+\frac{64 \zeta (5)}{3}
  \nonumber\\&&\mbox{}
-\frac{1168 \zeta (3)^2}{3}
-\frac{70208 \pi ^4 \zeta (3)}{945}
-\frac{3328 \pi ^2 \zeta (3)}{9}
-\frac{1504 \pi ^6}{135}
+\frac{14336 \pi ^4}{135}
+768 s_6\bigg) \epsilon ^2+O\left(\epsilon ^3\right)
  \,.
  \label{eq::I11}
\end{eqnarray} 
Note that the ${\cal O}(\epsilon^2)$ terms of $I_{16}$ and $I_{11}$ in
Eqs.~(\ref{eq::I16}) and~(\ref{eq::I11}) are not needed for $a_3^{[c]}$.
We nevertheless provide these results to demonstrate the powerfulness of the
DRA method.

In principle, the DRA method is also applicable to the calculation of
$I_{18}$. However, the difficulties related to the slow convergence of certain
matrix sums and the corresponding
precision loss appear to be overwhelming. For this reason, the method of
differential equations has been applied to $I_{18}$, see Ref.~\cite{LSS16}.

%- }}}
%- {{{ singlet:

\section{Singlet potential\label{sec::sing}}

In this Section we present analytic expressions for $a_3^{[1]}$. One- and two-loop
results using the same notation can be found in Ref.~\cite{Anzai:2013tja}.
Analytic results for the coefficients of $n_l^3$ and $n_l^2$ have already been
presented in Ref.~\cite{Smirnov:2008pn}. Here, they are repeated for
completeness
\begin{eqnarray}
  a_3^{[1],(3)} &=& -\left(\frac{20}9\right)^3 T_F^3 \,,\nonumber\\
  a_3^{[1],(2)} &=& \left(\frac{12541}{243} + \frac{368\zeta(3)}{3} 
    + \frac{64\pi^4}{135}\right) C_A T_F^2
  + \left(\frac{14002}{81} - \frac{416\zeta(3)}3\right) C_F T_F^2 \,.
\end{eqnarray}

Let us now turn to the $n_l^1$ and $n_l^0$ term.
Expressed in terms of the eigenvalues of the Casimir operators and higher
order group invariants $d_F^{abcd}$ and $d_A^{abcd}$ (see, e.g.,
Ref.~\cite{vanRitbergen:1998pn}) we obtain for
the linear-$n_l$ term the analytic result
\begin{eqnarray}
  a_3^{[1],(1)} &=& 
  \frac{d_F^{abcd}d_F^{abcd}}{N_A} \Bigg\{
     \pi^2\left(\frac{1264}{9} - \frac{976\zeta(3)}{3} + l_2\left(64 +
        672\zeta(3)\right)
      \right)
    + \pi^4\left(-\frac{184}{3} + \frac{32 l_2}{3} -   32 l_2^2\right)
    + \frac{10\pi^6}{3} 
  \Bigg\}
  \nonumber\\&&\mbox{}
  + T_F\Bigg\{
  C_F^2\left( \frac{286}{9} + \frac{296\zeta(3)}{3} - 160\zeta(5) \right)
  + C_A C_F \left( -\frac{71281}{162} + 264\zeta(3) + 80\zeta(5) \right) 
  \nonumber\\&&\mbox{}
  + C_A^2 \left[
    -\frac{58747}{486} 
    + \pi^2 \left( \frac{17}{27} - 32 \alpha_4
      + l_2 \left( -\frac{4}{3} - 14\zeta(3)\right) - \frac{19\zeta(3)}{3}
    \right)
    - 356\zeta(3) 
    \right.\nonumber\\&&\left.\mbox{}
    + \pi^4 \left(-\frac{157}{54} - \frac{5 l_2}{9} + l_2^2\right) 
    + \frac{1091\zeta(5)}{6}
    + \frac{57(\zeta(3))^2}{2} 
    + \frac{761\pi^6}{2520} 
    - 48s_6 
     \right]
     \Bigg\}
     \,,
\end{eqnarray}
and the gluonic part is given by
\begin{eqnarray}
  a_3^{[1],(0)} &=& 
  \frac{d_F^{abcd}d_A^{abcd}}{N_A}
  \Bigg\{
  \pi^2\left[ \frac{7432}{9}
    - 4736 \alpha_4 
    + l_2\left( \frac{14752}{3} - 3472\zeta(3)   \right) 
    - \frac{6616\zeta(3)}{3}
  \right]
  \nonumber\\&&\mbox{}
  + \pi^4\left(-156 + \frac{560 l_2}{3} + \frac{496 l_2^2}{3} \right)
  + \frac{1511\pi^6}{45} 
  \Bigg\}
  + C_A^3\Bigg\{
  \frac{385645}{2916} 
  + \pi^2 \left[
    - \frac{953}{54} + \frac{584 {\alpha_4} }{3}
    + \frac{175\zeta(3)}{2} 
    \right.\nonumber\\&&\left.\mbox{}
    + l_2 \left( -\frac{922}{9} +
      \frac{217\zeta(3)}{3} \right)
  \right]
    + \frac{584\zeta(3)}{3}
  + \pi^4\left( \frac{1349}{270} - \frac{20 l_2}{9} 
    - \frac{40 l_2^2}{9} 
  \right)
  - \frac{1927\zeta(5)}{6}
    - \frac{143(\zeta(3))^2}{2} 
  \nonumber\\&&\mbox{}
  - \frac{4621\pi^6}{3024} 
  + 144 s_6 
  \Bigg\}
  \,.
\end{eqnarray}
The numerical evaluation of the analytic results is in full agreement
(including all digits) with~\cite{Smirnov:2008pn,Smirnov:2009fh,Anzai:2009tm}.

It is interesting to note that the contributions proportional to
$d_F^{abcd}d_F^{abcd}$ and $d_F^{abcd}d_A^{abcd}$ only involve $\pi^2$,
$\pi^4$ and $\pi^6$ terms.  Note that these colour structures appear for the
first time at three-loop order.  On the other hand, the other colour
structures basically involve all constants one expects up to transcendentality
weight six.  Note, however, that the constant $s_6$ is only present in the
most non-abelian parts, i.e., $T_F C_A^2$ and $C_A^3$.  Let us also mention
that $\log(2)$ terms are present to first, second and fourth power but there
are no cubic terms.

In a next step we specify to SU($N_c$) and replace the colour factors
by
\begin{eqnarray}
  && C_A = N_c \,, \quad
  C_F = \frac{N_c^2-1}{2N_c} \,, \quad
  T_F = \frac{1}{2} \,, \quad 
  N_A=N_c^2-1\,,\nonumber\\&&
  \frac{d_F^{abcd}d_F^{abcd}}{N_A} = \frac{18 - 6N_c^2 + N_c^4}{96 N_c^2}\,, \quad
  \frac{d_F^{abcd}d_A^{abcd}}{N_A} = \frac{N_c(N_c^2+6)}{48}\,.
  \label{eq::ca2nc}
\end{eqnarray}
This leads to
\begin{eqnarray}
  a_3^{[1],(1)} &=& 
  \frac{66133}{648}
  + \pi^2 \left(-\frac{79}{9} + l_2 \left(-4 - 42\zeta(3)\right)
    + \frac{61\zeta(3)}{3}
  \right)
  - \frac{272\zeta(3)}{3}
  + \pi^4\left(\frac{23}{6} - \frac{2 l_2}{3} + 2 l_2^2 \right) 
  + 20 \zeta(5)
  - \frac{5 \pi^6}{24}
  \nonumber\\&&\mbox{}
  + \frac{1}{N_c^2}\Bigg\{
  \frac{143}{36} 
  + \pi^2 \left[\frac{79}{3} - 61\zeta(3) + l_2 \left( 12 + 126\zeta(3)\right)
  \right]
  + \frac{ 37 \zeta(3)}{3}
  + \pi^4 \left(-\frac{23}{2} + 2 l_2 - 6 l_2^2 \right)
  - 20 \zeta(5)
  \nonumber\\&&\mbox{}
  + \frac{ 5 \pi^6}{8}
  \Bigg\}
  + N_c^2 \Bigg\{
  - \frac{323615}{1944} 
  +  \pi^2 \left( \frac{16}{9} - 16 {\alpha_4} 
    - \frac{59 \zeta(3)}{9} \right)
  - \frac{299\zeta(3)}{3}
  + \pi^4 \left( -\frac{113}{54} - \frac{l_2}{6} + \frac{l_2^2}{6} \right)
  \nonumber\\&&\mbox{}
  + \frac{1091\zeta(5)}{12}
  + \frac{57(\zeta(3))^2}{4} 
  + \frac{13\pi^6}{70}
  - 24 s_6
  \Bigg\}
  \,,
  \nonumber\\
  a_3^{[1],(0)} &=& 
  N_c \Bigg\{
   \pi^2 \left[ \frac{929}{9} 
     - 592 {\alpha_4}
     + l_2 \left( \frac{1844}{3} - 434\zeta(3) \right) 
     - \frac{827\zeta(3)}{3}
   \right]
   + \pi^4 \left(-\frac{39}{2} + \frac{70 l_2}{3} + \frac{ 62 l_2^2 }{3}\right)
   + \frac{1511\pi^6}{360 }
  \Bigg\}
  \nonumber\\&&\mbox{}
  +  N_c^3 \Bigg\{
  \frac{385645}{2916} 
  + \pi^2 \left(-\frac{4}{9} + 96 {\alpha_4} + \frac{374\zeta(3)}{9}\right)
  + \frac{584 \zeta(3)}{3}
  + \pi^4 \left( \frac{943}{540} + \frac{5 l_2}{3} - l_2^2 \right) 
  \nonumber\\&&\mbox{}
  - \frac{1927 \zeta(5) }{6}
  - \frac{ 143 (\zeta(3))^2}{2}
  - \frac{29\pi^6}{35} 
  + 144 s_6
  \Bigg\}
  \,.
\end{eqnarray}

Finally, for $N_c=3$ we have
\begin{eqnarray}
  a_3^{[1],(1)} &=& 
  -\frac{452213}{324}
  +\pi^2\left[\frac{274}{27} - \frac{409\zeta(3)}{9} - 144 {\alpha_4}
    + l_2 \left(-\frac{8}{3} - 28\zeta(3)\right) 
  \right] 
  - \frac{26630\zeta(3)}{27} 
  \nonumber\\&&\mbox{}
  + \pi^4\left( -\frac{293}{18} - \frac{35 l_2}{18} 
    + \frac{17 l_2^2}{6} \right)
  + \frac{30097\zeta(5)}{36}
  + \frac{1931\pi^6}{1260} 
  + \frac{513(\zeta(3))^2}{4}
  - 216 s_6
  \,,
\end{eqnarray}
\begin{eqnarray}
  a_3^{[1],(0)} &=& 
  \frac{385645}{108} 
  + \pi^2\left[\frac{893}{3} + 816 {\alpha_4}
    + l_2 \left(1844 - 1302\zeta(3)\right) 
    + 295\zeta(3)
  \right] 
  + 5256\zeta(3)
  \nonumber\\&&\mbox{}
  + \pi^4 \left(-\frac{227}{20} + 115 l_2 + 35 l_2^2 \right) 
  - \frac{17343\zeta(5)}{2}
  - \frac{1643 \pi^6}{168} 
  - \frac{3861 (\zeta(3))^2}{2} 
  + 3888 s_6
  \,,
\end{eqnarray}
which in numerical form is given by
\begin{equation}
  a_3^{[1]} =
  13432.5648565 - 3289.9052968\, n_l
  +  185.9900266\, n_l^2 - 1.3717421\, n_l^3
  \,.
\end{equation}

%- }}}
%- {{{ octet:

\section{Octet potential\label{sec::oct}}

In this Section we proceed similar to the previous one and present results
for $\delta a_3^{[8],(i)}$ defined in Eq.~(\ref{eq::delta_a3}). We 
discuss the results in terms of $C_A$, $C_F$, etc. in
Appendix~\ref{app::a3o} and present in this section 
expressions in terms of $N_c$.
We have $\delta a_3^{[8],(i)} = 0$ for $i=2$ and $i=3$ and
for the linear-$n_l$ and $n_l$-independent terms we get
\begin{eqnarray}
  \delta a_3^{[8],(1)} &=& 
   \pi^2 \left[-\frac{11}{3} - 31\zeta(3) + l_2 \left(4 + 42\zeta(3)\right)  \right]
  + \pi^4 \left(-\frac{7}{6} + \frac{2 l_2}{3} - 2 l_2^2 \right)
  + \frac{5\pi^6}{24} 
  \nonumber\\&&\mbox{}
  + N_c^2 \Bigg[
    \pi^2\left(\frac{8}{9} + 48 {\alpha_4} + 25 \zeta(3) \right)
  + \pi^4\left(\frac{2}{3} + \frac{2 l_2}{3}\right)
  - \frac{13\pi^6}{20} 
  \Bigg]
  \,,
  \nonumber\\
  \delta a_3^{[8],(0)} &=& 
  N_c^3\Bigg\{
  \pi^2 \left[ \frac{139}{9} + 304 {\alpha_4} + 15 \zeta(3) 
    + l_2 \left( -\frac{1844}{3} + 434 \zeta(3)\right)
  \right]
  \nonumber\\&&\mbox{}
  + \pi^4 \left( \frac{295}{6} - 30 l_2 - \frac{62 l_2^2}{3} \right) 
  -\frac{1187\pi^6}{360} 
  \Bigg\}
  \,,
\end{eqnarray}
which for $N_c=3$ leads to
\begin{eqnarray}
  \delta a_3^{[8],(1)} &=& 
  - \frac{677\pi^6}{120} 
  + \pi^4 \left( \frac{29}{6} + \frac{20 l_2}{3} - 2 l_2^2 \right) 
  + \pi^2 \left[ \frac{13}{3} + 432 {\alpha_4} + 194 \zeta(3) 
    + l_2 \left( 4 + 42 \zeta(3) \right)
  \right]
  \,,
  \nonumber\\
  \delta a_3^{[8],(0)} &=& 
  \pi^2\left[417 + 8208 {\alpha_4} + 405 \zeta(3) +
    l_2 \left( -16596 + 11718 \zeta(3) \right)
  \right]
  + \pi^4\left(\frac{2655}{2} - 810 l_2 - 558 l_2^2\right)
  \nonumber\\&&\mbox{}
  - \frac{3561\pi^6}{40} 
  \,.
\end{eqnarray}

It is interesting to note that $\delta a_3^{[8],(0)}$ and $\delta
a_3^{[8],(1)}$ have an overall factor $\pi^2$ which was predicted in
Ref.~\cite{Anzai:2013tja} on the basis of the involved master
integrals. Although they could not be computed analytically it was possible to
show that there is an overall factor $\pi^2$, a feature which is
also observed at two-loop order in QCD~\cite{Kniehl:2004rk,Collet:2011kq} and
in ${\cal N}=4$ supersymmetric Yang Mills theories~\cite{Prausa:2013qva}.

In numerical form we obtain for the complete three-loop coefficient
\begin{eqnarray}
  \delta a_3^{[8]} &=&
  -2634.7351731 + 367.9626044 \, n_l
  \,.
\end{eqnarray}

%- }}}
%- {{{ Concl.:

\section{Conclusions\label{sec::concl}}

The interaction of a slowly moving heavy quark-anti-quark pair can be described
with the help of a static potential, a concept which is familiar from ordinary
quantum mechanics. Its perturbative part is obtained from the exchange of
soft gluons which are conveniently considered in the framework of
non-relativistic QCD. Numerical results for the three-loop potential, which
have entered a number of physical observables, have been
obtained eight years ago by two independent
groups~\cite{Smirnov:2008pn,Smirnov:2009fh,Anzai:2009tm}.  The obtained
precision has been sufficient for all physical applications where $a_3$
entered as a building block. However, from the aesthetic point of view it is
important to obtain analytic results for higher order quantum corrections.
This has been achieved in this paper.  We have obtained analytic
results for the three-loop corrections to the singlet and octet potential
which are presented in Sections~\ref{sec::sing} and~\ref{sec::oct}, respectively.

%- }}}

%- {{{ Ackn.:

%\medskip

\section*{Acknowledgements}
We thank Alexander Penin for carefully reading the manuscript.
R.L. acknowledges support through RFBR grant No. 15-02-07893.

%- }}}

%- {{{ Appendix:

\begin{appendix}

\section{\label{app::a3o}$\delta a_2^{[8]}$ and $\delta a_3^{[8]}$ 
  in terms of colour invariants}

In this appendix we present results for $\delta a_3^{[8],(1)}$ and $\delta
a_3^{[8],(0)}$ in terms of $C_A$, $C_F$, $T_F$, $N_A$, $d_F^{abcd}$ and
$d_A^{abcd}$. Let us mention that the representation given in
Eq.~(\ref{eq::V}) is only valid for SU$(N_c)$. Thus, in the following
we present results for $C^{[8]} \delta a_3^{[8],(i)}$ ($i=0,1$) with
$C^{[8]}=C_F-C_A/2$. For completeness we also present the two-loop
expression; at one-loop order we have $\delta a_1^{[8]}=0$. Our results read
\begin{eqnarray}
  C^{[8]} \delta a_2^{[8]} &=& 
  \left(\frac{\pi^4}{12}-\pi^2\right)
  \left(C_A^3 - 48 \frac{d_F^{abcd}d_A^{abcd}}{N_A} \right)
  \,,\nonumber\\
  C^{[8]} \delta a_3^{[8],(1)} &=& 
  C_A \frac{d_F^{abcd}d_F^{abcd}}{N_A} \left[
     \pi^2\left(\frac{88}{9} - \frac{32l_2}{3} + \frac{248\zeta(3)}{3} 
       - 112\zeta(3)l_2\right)
    + \pi^4\left(\frac{28}{9} - \frac{16 l_2}{9} + \frac{16 l_2^2}{3}\right)
    -\frac{5\pi^6}{9}
  \right]
  \nonumber\\&&\mbox{}
  + \frac{d_F^{abcd}d_A^{abcd}}{N_A} \left[
      \pi^2 \left(\frac{4}{3} - 192\alpha_4 - \frac{16 l_2}{3} 
        - \frac{176\zeta(3)}{3} - 56 l_2 \zeta(3) \right)
    + \pi^4 \left(-\frac{10}{9} - \frac{32 l_2}{9} + \frac{8 l_2^2}{3}\right)
    + \frac{209 \pi^6}{90}
    \right]
  \nonumber\\&&\mbox{}
  + C_A^3 T_F \left[
      \pi^2 \left(-\frac{7}{27} + 8\alpha_4 + \frac{4 l_2}{9} 
        + \frac{13 \zeta(3)}{18} + \frac{14 l_2 \zeta(3)}{3} \right)
    + \pi^4 \left(-\frac{1}{54} + \frac{5 l_2}{27} - \frac{2 l_2^2}{9}\right)
    - \frac{23 \pi^6}{270}
    \right]
  \,,\nonumber\\
  C^{[8]} \delta a_3^{[8],(0)} &=&
  C_A \frac{d_F^{abcd}d_A^{abcd}}{N_A} \left[
     \pi^2\left(-\frac{2356}{9} + 3520\alpha_4 - \frac{7376 l_2}{3} + 1420\zeta(3) 
       + 1736\zeta(3)l_2\right)
    + \pi^4\left( 66 - \frac{200 l_2}{3} - \frac{248 l_2^2}{3}\right)
    \right.\nonumber\\&&\left.\mbox{}
    -\frac{511\pi^6}{18}
  \right]
  + \frac{d_A^{abcd}d_A^{abcd}}{N_A} \left[
      \pi^2 \left(\frac{50}{3} - \frac{1184\alpha_4}{3} + \frac{3688 l_2}{9} 
        - \frac{370\zeta(3)}{3} - \frac{868 l_2 \zeta(3)}{3} \right)
    \right.\nonumber\\&&\left.\mbox{}
    + \pi^4 \left(-\frac{197}{9} + \frac{140 l_2}{9} + \frac{124 l_2^2}{9}\right)
    + \frac{1871 \pi^6}{540}
    \right]
  + C_A^4 \left[
      \pi^2 \left(\frac{257}{54} - \frac{512\alpha_4}{9} + \frac{922 l_2}{27} 
        - \frac{220 \zeta(3)}{9} 
    \right.\right.\nonumber\\&&\left.\left.\mbox{}
        - \frac{217 l_2 \zeta(3)}{9} \right)
    + \pi^4 \left(-\frac{25}{54} + \frac{20 l_2}{27} + \frac{31 l_2^2}{27}\right)
    + \frac{2897 \pi^6}{6480}
    \right]
  \,,
  \label{eq::delta_a3_oct_gen}
\end{eqnarray}
with
\begin{eqnarray}
  \frac{d_A^{abcd}d_A^{abcd}}{N_A} = \frac{N_c^2(N_c^2+36)}{24}\,.
\end{eqnarray}
Numerical results of Eq.~(\ref{eq::delta_a3_oct_gen}) are given in
Ref.~\cite{prausa_dipl}.  All colour factors have been computed with the help
of the program {\tt color}~\cite{vanRitbergen:1998pn}.

\end{appendix}

%- }}}

%- {{{ Refs.:

%- }}}

\end{document}